\documentclass[onecolumn,showpacs]{revtex4}

\usepackage{graphicx}% Include figure files
\usepackage{dcolumn}% Align table columns on decimal point
\usepackage{bm}% bold math

\input epsf

\begin{document}

\title{\Large Role of Modified Chaplygin Gas as a Dark Energy Model in Collapsing Spherically Symmetric Cloud}

\author{\bf Ujjal
Debnath$^1$\footnote{ujjaldebnath@yahoo.com} and Subenoy
Chakraborty$^2$\footnote{subenoyc@yahoo.co.in}}

\affiliation{$^1$Department of Mathematics, Bengal Engineering and
Science University, Shibpur, Howrah-711 103, India.\\
$^2$Department of Mathematics, Jadavpur University, Calcutta-32,
India.\\ }

\date{\today}

\begin{abstract}
In this work gravitational collapse of a spherical cloud, consists
of both dark matter and dark energy in the form of modified
Chaplygin gas is studied. It is found that dark energy alone in
the form of modified Chaplygin gas forms black hole. Also when
both components of the fluid are present then the collapse favors
the formation of black hole in cases the dark energy dominates
over dark matter. The conclusion is totally opposite to the
usually known results.
\end{abstract}


\maketitle

\section{\normalsize\bf{Introduction}}

Recent observational data [1, 2] shows the consistency of the
inflationary scenario with the power spectrum of the microwave
background radiation for cosmic fluid having equation of state in
the range $-1\le \gamma(=p/\rho)\le -1/3$. to match with these
observational results usually two (dark) components of matter are
invoked : the pressureless cold dark matter (or simply dark matter
(DM)) and the dark energy (DE) having negative pressure
components. The DM contribution $(\Omega_{DM}\sim 0.3)$ is mainly
motivated by the theoretical study of galactic rotation curves and
large scale structure formation, while for dark energy
$\Omega_{DE}\sim 0.7$ and is responsible for the acceleration of
the distant type Ia supernovae (for recent reviews see [3] and
[4]). Though there are no direct laboratory observational or
experimental evidence for both of them, yet a unified dark
matter-dark energy scenario i.e., they are two different
manifestations of a single fluid [5] would be interesting.
Recently, unified model has been proposed which is known as {\it
modified Chaplygin gas} [6, 7] having exotic equation of state

\begin{equation}
p=\gamma \rho-\frac{B}{\rho^{\alpha}}~,~~~B>0,~~~0<\alpha<1
\end{equation}

In this paper, gravitational collapse of a spherically symmetric
cloud consists of both dark matter and dark energy (having
equation of state given by equation (1)) is considered with
energy-momentum tensor

\begin{equation}
T^{j}_{i}=(\rho_{DM}+\rho+p)u_{i}u^{j}-p\delta^{j}_{i}
\end{equation}

The Einstein equations for spherical space-time with line-element

\begin{equation}
ds^{2}=dt^{2}-a^{2}(t)\left(dr^{2}+r^{2}d\Omega^{2} \right)
\end{equation}

are given by

\begin{equation}
3\frac{\dot{a}^{2}}{a^{2}}=\kappa(\rho_{DM}+\rho)
\end{equation}
and
\begin{equation}
2\frac{\ddot{a}}{a}+\frac{\dot{a}^{2}}{a^{2}}=\kappa\rho
\end{equation}

Now, if $Q(t)$ denotes the interaction between dark matter and
dark energy then from the conservation law $T^{\nu}_{\mu;\nu}=0$
one gets

\begin{equation}
\dot{\rho}_{DM}+3\frac{\dot{a}}{a}\rho_{DM}=Q
\end{equation}
and
\begin{equation}
\dot{\rho}+3\frac{\dot{a}}{a}(\rho+p)=-Q
\end{equation}

If $\Sigma~:~r=r_{\Sigma}$ denotes the boundary of the spherical
collapsing cloud then on $\Sigma$

\begin{equation}
ds^{2}_{\Sigma}=dT^{2}-R^{2}(T)d\Omega^{2}
\end{equation}

where $T=t$ and $R(T)=r_{\Sigma}a(T)$ is called the area radius.
Thus the total mass of the collapsing cloud is

\begin{equation}
M(T)=\frac{1}{2}R(T)\dot{R}^{2}(T)
\end{equation}

If $T_{AH}$ be the time instant at which the whole cloud starts to
be trapped then

\begin{equation}
R_{,\alpha}R_{,\beta}g^{\alpha\beta}|_{T=T_{AH}}=0~~,~~~~{\text
i.e.,} ~~\dot{R}^{2}(T_{AH})=1
\end{equation}

As it is natural to assume the cloud to be untrapped initially
($t=t_{i}$) so one should have

\begin{equation}
\dot{R}(T=T_{i})>-1
\end{equation}

Note that if the condition (11) holds throughout the collapsing
process then the collapse will not produce black holes. in the
following two sections, collapsing process will be studied when
there is only Chaplygin gas as the collapsing fluid and then a
combination of dark matter and Chaplygin gas both with and without
interaction. the paper ends with some conclusive remarks.\\

\section{\normalsize\bf{Gravitational Collapse of  Dark Energy  as Chaplygin gas model}}

This section deals with gravitational collapse of dark energy in
the form of Chaplygin gas. From the conservation equation (7),
integrating once one gets

\begin{equation}
\rho=\left[\frac{B}{1+\gamma}+\frac{C}{a^{3(1+\gamma)(1+\alpha)}}
\right]^{\frac{1}{1+\alpha}}~,~~~(\gamma\ne -1)
\end{equation}

with $C$ is the constant of integration.\\

Now substituting this expression for $\rho$ into the Friedman
equation (4) and integrating the scale factor can be obtained as

\begin{equation}
a^{\frac{3(1+\gamma)}{2}}~_{2}F_{1}[x,x,1+x,-\frac{B}{C(1+\gamma)}a^{\frac{3(1+\gamma)}{2x}}]
=\frac{\sqrt{3\kappa}}{2}(1+\gamma)C^{x}(t_{0}-t)
\end{equation}

where $x=\frac{1}{2(1+\alpha)}$ and $_{2}F_{1}$ is the
hypergeometric function.\\

The expressions for the related physical parameters are

\begin{equation}
\dot{R}(\tau)=-R_{0}~a^{-\frac{3(1+\gamma)}{2}}\left[C+\frac{B}{1+\gamma}~a^{3(1+\alpha)(1+\gamma)}
\right]^{\frac{1}{2(1+\alpha)}}
\end{equation}

\begin{equation}
M(\tau)=\frac{1}{2}~R_{0}^{2}~r_{\Sigma}~a^{-3\gamma}\left[C+\frac{B}{1+\gamma}~a^{3(1+\alpha)(1+\gamma)}
\right]^{\frac{1}{(1+\alpha)}}
\end{equation}

One may note that as $t\rightarrow t_{0}$

$$
a^{\frac{3(1+\gamma)}{2}}\simeq
\frac{\sqrt{3\kappa}}{2}(1+\gamma)C^{\frac{1}{2(1+\alpha)}}(t_{0}-t)\sim
0
$$

Also using the relation [8]

\begin{equation}
_{2}F_{1}[a,b,c;z]=\frac{\Gamma(c)\Gamma(b-a)}{\Gamma(b)\Gamma(c-a)}~(-z)^{-a}~_{2}F_{1}[a,1-c+a,1-b+a;\frac{1}{z}]+
\frac{\Gamma(c)\Gamma(a-b)}{\Gamma(a)\Gamma(c-b)}~(-z)^{-b}~_{2}F_{1}[b,1-c+b,1-a+b;\frac{1}{z}]
\end{equation}

one gets the limiting value of

$$
a^{\frac{3(1+\gamma)}{2}}~_{2}F_{1}[\frac{1}{2(1+\alpha)},\frac{1}{2(1+\alpha)},1+\frac{1}{2(1+\alpha)},
-\frac{B}{C(1+\gamma)}a^{3(1+\alpha)(1+\gamma)}]
$$

as

$$
\frac{1}{1+\alpha}\left[\frac{C(1+\gamma)}{B}
\right]^{\frac{1}{2(1+\alpha)}}
$$
when $a$ is very large. thus if $t\rightarrow t_{s}$ as
$a\rightarrow\infty$ then from equation (13)

\begin{equation}
t_{s}=t_{0}-\frac{2}{\sqrt{3\kappa}~(1+\alpha)(1+\gamma)}\left(\frac{1+\gamma}{B}
\right)^{\frac{1}{2(1+\alpha)}}
\end{equation}

The limiting value of the physical parameters are

$$
\tau\rightarrow\tau_{s}~~:~~\rho\rightarrow
\left[\frac{B}{1+\gamma}\right]^{\frac{1}{1+\alpha}}~,~~~\dot{R}\rightarrow
\left\{\begin{array}{lll} -\infty ~~\text{for}~~\gamma>-5/3\\\\
~~0~~~~\text{for}~~\gamma\le -5/3
\end{array}\right.~ ,~~ M(\tau)\rightarrow \infty
$$
$$
\tau\rightarrow\tau_{0}~~:~~\rho\rightarrow
\infty~,~~\dot{R}\rightarrow -\infty~,~~M(\tau)\rightarrow
\infty~~~~~~~~~~~~~~~~~~~~~~~~~~~~~~~~~~~~~~~~
$$

Thus if the collapse starts at an instant close to $\tau_{s}$ then
for $\gamma>-5/3$, initially the collapsing system is trapped and
in course of the collapsing process it gets untrapped (provided
the maximum value of $\dot{R}$ is greater than $-1$) and then
again it is trapped and black hole forms. However, for $\gamma\le
-5/3$, the system is initially untrapped and as it approaches to
the singularity at $\tau=\tau_{0}$, it gets trapped and leads to
the formation of a black hole. Thus dark energy alone in the form
of Chaplygin gas favours formation of black hole.\\

\section{\normalsize\bf{Collapsing process under the joint influence of dark matter and dark energy}}

This section is divided into two parts. In the first case, the
interaction $Q(t)$ is neglected while in the second case, the
influence of $Q(t)$ is considered.\\

{\bf CASE I : Interaction is neglected : $Q(t)=0$:}\\

Here  the conservation equation for $\rho_{DM}$ gives

\begin{equation}
\rho_{DM}=\frac{\rho_{0}}{a^{3}}~~,~~~\rho_{0}>0~~~~\text{a~
constant}.
\end{equation}

The expressions for $\dot{R}(\tau)$ and $M(\tau)$ are

\begin{equation}
\dot{R}(\tau)=-R_{0}~a^{\frac{1}{2}}\left[\rho_{0}+a^{-3\gamma}\left\{C+\frac{B}{1+\gamma}a^{3(1+\alpha)
(1+\gamma)} \right\}^{\frac{1}{1+\alpha}} \right]^{\frac{1}{2}}
\end{equation}
and
\begin{equation}
M(\tau)=\frac{1}{2}~R_{0}^{2}~r_{\Sigma}~a^{2}\left[\rho_{0}+a^{-3\gamma}\left\{C+\frac{B}{1+\gamma}a^{3(1+\alpha)
(1+\gamma)} \right\}^{\frac{1}{1+\alpha}} \right]
\end{equation}

with $R_{0}=r_{\Sigma}~\sqrt{\frac{\kappa}{3}}$.\\

As the integral in equation (19) can not be evaluated in general,
so only the approximate forms for `$a$' may be obtained for small
and large `$a$'. However, one can determine the behaviour of the
physical parameters in these two limits (namely, $a\rightarrow 0$
and $a\rightarrow \infty$) as follows:

\begin{eqnarray*}
a\rightarrow 0~:~~\rho_{DM}\rightarrow \infty~,~~\rho\rightarrow
\left\{\begin{array}{lll} \infty~~if~~1+\gamma>0\\\\a~
constant~~if~~1+\gamma<0
\end{array}\right.~~,~~
\dot{R}\rightarrow \left\{\begin{array}{lll} 0, ~~if
~~\gamma<\frac{1}{3}\\\\
-\infty,~~if~~\gamma>\frac{1}{3}\\\\
-\mu~,~if~~\gamma=\frac{1}{3}
\end{array}\right.~~,
\end{eqnarray*}

\begin{equation}
M\rightarrow \left\{\begin{array}{lll}
0,~~if~~\gamma<\frac{2}{3}\\\\
\infty,~~if~~\gamma>\frac{2}{3}\\\\
a~constant,~~if~~\gamma=\frac{2}{3}
\end{array}\right.,~~~\mu=R_{0}~C^{\frac{1}{2(1+\alpha)}}.
\end{equation}

\begin{equation}
a\rightarrow \infty~:~~\rho_{DM}\rightarrow 0,~~\rho\rightarrow
\left\{\begin{array}{lll} a~constant,~~if~~1+\gamma>0\\\\
\infty,~~if~~1+\gamma<0
\end{array}\right.,~~\dot{R}\rightarrow -\infty,~~M\rightarrow
\infty.~~~~~~~~~~~~~~~~~~~~
\end{equation}

Thus $a=0$ is always a singularity of the space-time and it is
covered by an apparent horizon for $\gamma\ge 1/3$ (provided
$\mu>1$) while the singularity is naked for $\gamma<1/3$. n the
otherhand, $a=\infty$ may be singular if $\gamma<-1$ and always
black hole will form.\\

{\bf CASE II : Gravitational Collapse with Interaction:}\\

Recently, Cai and Wang [9, 10] have assumed the ratio of dark
energy density and dark matter density as

\begin{equation}
\frac{\rho}{\rho_{DM}}=A~a^{3n}
\end{equation}

with $A>0$ and $n$ as arbitrary constants. then solving the
conservation equations (6) and (7) one obtains

\begin{eqnarray*}
\rho_{t}^{\alpha+1}=\frac{(\alpha+1)B}{[\alpha(n-1)-1]}\frac{\left(A~a^{3n}\right)^{\frac{2}{n}
(\alpha+1-n\alpha)-(\alpha+1)}}{\left(A~a^{3n}+1\right)^{\frac{2}{n}
(\alpha+1-n\alpha)+\gamma(\alpha+1)}}\times~~~~~~~~~~~~~~~~~~~~~~~~~~~~~~~~~~~~~
\end{eqnarray*}

\begin{eqnarray*}
_{2}F_{1}[\frac{1+\alpha-n\alpha}{n},\frac{1+n+\alpha+\gamma+\alpha\gamma}{n},\frac{1+n+\alpha-n\alpha}{n},
\frac{A~a^{3n}}{1+A~a^{3n}}]
\end{eqnarray*}

\begin{equation}
+z_{0}~\left[A~a^{3n}\left(1+A~a^{3n}\right)^{\gamma}\right]^{-(\alpha+1)}
\end{equation}

where $\rho_{t}=\rho+\rho_{DM}$ and using (23) one gets

\begin{equation}
\rho=\frac{A~a^{3n}~\rho_{t}}{1+A~a^{3n}},~~\rho_{DM}=\frac{\rho_{t}}{1+A~a^{3n}}
\end{equation}

Hence from the conservation equation (6) and the Friedman
equation (4) the expression for the interaction is

\begin{equation}
Q(t)=-\frac{3(\gamma+n)A~a^{3n}~\rho_{t}}{(1+A~a^{3n})^{2}}~\frac{\dot{a}}{a}+\frac{3B(1+A~a^{3n})^{\alpha-1}}
{\rho_{t}^{\alpha}A^{\alpha}a^{3n\alpha}}~\frac{\dot{a}}{a}
\end{equation}

where
\begin{eqnarray*}
\frac{\dot{a}}{a}=-\frac{\rho_{0}}{a^{\frac{3n}{2}}(1+A~a^{3n})^{\frac{\gamma}{2}}}
\left[\frac{a^{6(\alpha+1-n\alpha)}A^{\frac{2}{n}(\alpha+1-n\alpha)}}{(1+A~a^{3n})^{\frac{2}{n}(\alpha+1-n\alpha)}}~\times
~~~~~~~~~~~~~~~~~~~~~~~~~~~~~~~~~~~~~~~~~\right.
\end{eqnarray*}
\begin{equation}
 \left.
_{2}F_{1}[\frac{1+\alpha-n\alpha}{n},\frac{1+n+\alpha+\gamma+\alpha\gamma}{n},\frac{1+n+\alpha-n\alpha}{n},
\frac{A~a^{3n}}{1+A~a^{3n}}] +z_{1}
\right]^{\frac{1}{2(\alpha+1)}}
\end{equation}
with
$$
\rho_{0}=\sqrt{\frac{\kappa}{3A}}~\left[\frac{(\alpha+1)B}{\alpha(n-1)-1}
\right]^{\frac{1}{2(\alpha+1)}}~,~~~~z_{1}=z_{0}~\left[\frac{(\alpha+1)B}{\alpha(n-1)-1}
\right]^{-\frac{1}{2(\alpha+1)}}
$$

The equation (27) can be written in the integral form as

\begin{equation}
\int
\frac{\left(1+Ay^{2}\right)^{\frac{\gamma}{2}}dy}{\left[\frac{(Ay^{2})^{\frac{2}{n}(\alpha+1-n\alpha)}}
{(1+A~y^{2})^{\frac{2}{n}(\alpha+1-n\alpha)}}~_{2}F_{1}[\frac{1+\alpha-n\alpha}{n},\frac{1+n+\alpha+\gamma+
\alpha\gamma}{n},\frac{1+n+\alpha-n\alpha}{n},\frac{A~y^{2}}{1+A~y^{2}}]
+z_{1} \right]^{\frac{1}{2(\alpha+1)}}}=-y_{0}(t-t_{0})
\end{equation}

with $y=a^{3n/2}$ and $y_{0}=\frac{3n}{2}~\rho_{0}$.\\

The expression for mass function and $\dot{R}$ are

\begin{eqnarray*}
M(\tau)=\frac{r_{\Sigma}^{3}~\rho_{0}^{2}}{2a^{3n-1}\left(1+A~a^{3n}\right)^{\gamma}}~
\left[\frac{a^{6(\alpha+1-n\alpha)}A^{\frac{2}{n}(\alpha+1-n\alpha)}}{(1+A~a^{3n})^{\frac{2}{n}(\alpha+1-n\alpha)}}~\times
~~~~~~~~~~~~~~~~~~~~~~~~~~~~~~~~~~~~~~~~~~~~~~~~\right.
\end{eqnarray*}
\begin{equation}
 \left.
_{2}F_{1}[\frac{1+\alpha-n\alpha}{n},\frac{1+n+\alpha+\gamma+\alpha\gamma}{n},\frac{1+n+\alpha-n\alpha}{n},
\frac{A~a^{3n}}{1+A~a^{3n}}] +z_{1} \right]^{\frac{1}{\alpha+1}}
\end{equation}
and
\begin{eqnarray*}
\dot{R}(\tau)=-\frac{\rho_{0}~r_{\Sigma}}{a^{\left(\frac{3n}{2}-1\right)}\left(1+A~a^{3n}\right)^{\frac{\gamma}{2}}}~
\left[\frac{a^{6(\alpha+1-n\alpha)}A^{\frac{2}{n}(\alpha+1-n\alpha)}}{(1+A~a^{3n})^{\frac{2}{n}(\alpha+1-n\alpha)}}~\times
~~~~~~~~~~~~~~~~~~~~~~~~~~~~~~~~~~~~~~~~~\right.
\end{eqnarray*}
\begin{equation}
 \left.
_{2}F_{1}[\frac{1+\alpha-n\alpha}{n},\frac{1+n+\alpha+\gamma+\alpha\gamma}{n},\frac{1+n+\alpha-n\alpha}{n},
\frac{A~a^{3n}}{1+A~a^{3n}}] +z_{1}
\right]^{\frac{1}{2(\alpha+1)}}
\end{equation}

The above expressions for the physical parameters show the
following limiting behaviour

$$
\rho_{t}\sim \left\{\begin{array}{lll} a^{-3n},~~a\rightarrow
0\\\\
a^{-3n(\alpha+1)(\gamma+1)},~~a\rightarrow \infty
\end{array}\right.~,~~ \rho \sim \left\{\begin{array}{lll}
a~constant,~~a\rightarrow
0\\\\a^{-3n(\alpha+1)(\gamma+1)},~~a\rightarrow\infty
\end{array}\right.~,
$$
$$
\rho_{DM}\sim \left\{\begin{array}{lll} a^{-3n},~~a\rightarrow
0\\\\
a^{-3n(\alpha+1)(\gamma+1)},~~a\rightarrow \infty
\end{array}\right.~,~~ \dot{R}(\tau) \sim \left\{\begin{array}{lll}
-a^{1-\frac{3n}{2}},~~a\rightarrow
0\\\\-a^{1-\frac{3n}{2}(1+\gamma)},~~a\rightarrow\infty
\end{array}\right.~,
$$
$$
M(\tau)\sim \left\{\begin{array}{lll} a^{1-3n},~~a\rightarrow
0\\\\
a^{1-3n(\gamma+1)},~~a\rightarrow \infty
\end{array}\right.
$$
It is to ne noted that $a=0$ is always a singularity of the
space-time but $a=\infty$ is singularity if $1+\gamma<0$.\\

The above integral in equation (28) is solvable for the (choice
$\alpha=1$, $n=2$ and one gets) restriction $1+\alpha=n\alpha$ and
one gets

\begin{equation}
\frac{y}{(1+z_{1})^{\frac{1}{4}}}~_{2}F_{1}[\frac{1}{2},-\frac{\gamma}{2},\frac{3}{2},-Ay^{2}]=-y_{0}(t-t_{0})
\end{equation}

Thus in the limit as $a\rightarrow 0$, one finds

$$
a\sim
a_{0}(t_{0}-t)^{\frac{1}{3}}~,~~a_{0}=y_{0}^{\frac{2}{3}}(1+z_{1})^{\frac{1}{6}}~~~i.e.,~~a\sim
0~~as~~t\rightarrow t_{0}~.
$$

Further, using the property of the hypergeometric function (see
equation (16)), for large $a$, the solution (31) approximates to

$$
a^{1+\gamma}\sim
a_{1}(t_{0}-t)~,~~a_{1}=\frac{4a_{0}}{(1+\gamma)A^{\frac{\gamma}{2}}}~,~~for~~\gamma>-1
$$
and
$$
\frac{\sqrt{\pi}~~\Gamma(-\frac{\gamma+1}{2})}{2\sqrt{A}~(1+z_{1})^{\frac{1}{4}}~~\Gamma(-\frac{\gamma}{2})}
=-y_{0}(t_{s}-t_{0})
$$
$$
i.e.,~~~~t_{s}=t_{0}-\frac{\sqrt{\pi}~~\Gamma(-\frac{\gamma+1}{2})}{2\sqrt{A}~a_{0}^{3}
~~\Gamma(-\frac{\gamma}{2})}~,~~~for~~\gamma\le -1
$$

The limiting value of the physical parameters show that if $n<2/3$
then the space-time collapses to a naked singularity while black
hole will form for $n>2/3$. However, the singularity at $a=\infty$
always corresponds to a black hole solution for $1+\gamma<0$.\\

\section{\normalsize\bf{Conclusion}}

The paper deals with gravitational collapse of a spherically
symmetric homogeneous and isotropic fluid having finite radius.
The fluid has two component $-$ one component is dark matter in
the form os dust and the other, the dark energy component is the
modified Chaplygin gas model.\\

When the collapsing fluid is only in the form of modified
Chaplygin gas (the dark energy) then the collapse always leads to
the formation of a black hole. But there is some peculiarity for
$\gamma>-5/3$. Initially, the space-time is trapped and during the
evolution it gets untrapped and again it is covered by an apparent
horizon. This feature is interpreted by Cai and wang [10] (see
also [11]) as the evaporation of a white hole by ejecting matter
which again re-collapse to form a black hole. Note that the
collapsing dark energy in the form of Chaplygin gas can alone form
black holes unlike the dark energy model of Cai and wang [10] is
not in favour of black hole.\\

Section III deals with collapsing fluid having both components
with or without interaction. In both cases it is also found that
when the dark energy density dominates over the dark matter energy
density then the collapse favours formation of black hole.
Further, the expression for the interaction parameter has two
terms of which the first one is identical to that of Cai and wang
[10] while the second term, due to the Chaplygin gas having
negative sign reduces the interaction parameter. Therefore, from
the above study, one may conclude that the dark energy is not
always against the formation of black holes, it favours the
formation of apparent horizon in some cases.\\

{\bf Acknowledgement:}\\

One of the authors (SC) is thankful to CSIR, Govt. of India for
providing a research project No. 25(0141)/05/EMR-II.\\

{\bf References:}\\
\\
$[1]$ S. Perlmutter et al, {\it Astrophys. J.} {\bf 517} 565
(1998); A. G. Riess et al, {\it Astrophys. J.} {\bf 116} 109 (1998).\\
$[2]$ P. de Bernardis et al, {\it Nature} {\bf 404} 995 (2000); S.
Hanany et al, {\it Astrophys. J.} {\bf 545} L5 (2000).\\
$[3]$ P. J. E. Peebles and B. Ratra, {\it Rev. Mod. Phys.} {\bf
75} 559 (2003).\\
$[4]$ T. Padmanabhan, {\it Phys. Reports} {\bf 380} 235 (2003).\\
$[5]$ T. Matos and A. Urena-Lopez, {\it Class. quantum Grav.} {\bf
17} L75 (2000); C. Wetterich, {\it Phys. Rev. D} {\bf 65} 123512
(2002); T. Padmanabhan and T. R. Choudhury, {\it Phys. Rev. D}
{\bf 66} 081301 (2002).\\
$[6]$ H. B. Benaoum, {\it hep-th}/0205140.\\
$[7]$ U. Debnath, A. Banerjee and S. Chakraborty, {\it Class.
Quantum Grav.} {\bf 21} 5609 (2004).\\
$[8]$ M Abramowitz and I. A. Stegun, {\it Handbook of Mathematical
Functions}, (Dover Publications, INC, New York, 1972), pp. 555 -
566.\\
$[9]$ R. -G. Cai and A. Wang, {\it JCAP} {\bf 0503} 002 (2005).\\
$[10]$ R. -G. Cai and A. Wang, {\it astro-ph}/0505136.\\
$[11]$ D. M. Eardly, {\it Phys. Rev. Lett.} {\bf 33} 442 (1974);
S. K. Blau, {\it Phys. Rev. D} {\bf 39} 2901 (1989).\\

\end{document}